\documentclass{jpsj3}
\usepackage{txfonts}
\title{Large Zeeman Splitting in Out-of-Plane Magnetic Field in a Double-Layer Quantum Point Contact}

\author{Daiju Terasawa$^1$, Shota Norimoto$^2$, Tomonori Arakawa$^{2,3}$, Meydi Ferrier$^{2,4}$, Akira Fukuda$^1$, Kensuke Kobayashi$^{2,5}$, and Yoshiro Hirayama$^6$}
\inst{$^1$Department of Physics, Hyogo College of Medicine, Nishinomiya 663-8501, Japan \\
$^2$Graduate School of Science, Department of Physics, Osaka University, Toyonaka 560-0043, Japan \\
$^3$Center for Spintronics Research Network, Osaka University, Toyonaka, Osaka 560-8531, Japan \\
$^4$Laboratoire de Physique des Solides, CNRS, Universit\'e Paris-Sud, Universit\'e Paris Saclay, 91405 Orsay Cedex, France \\
$^5$Institute for Physics of Intelligence and Department of Physics, The University of Tokyo, Tokyo 113-0033, Japan \\
$^6$Graduate School of Science and CSIS, Tohoku University, Sendai 980-8578, Japan} 

\abst{In this study, we observe that the conductance of a quantum point contact on a GaAs/AlGaAs double quantum well depends significantly on the magnetic field perpendicular to the two-dimensional electron gas.
In the presence of the magnetic field, the subband edge splitting due to the Zeeman energy reaches 0.09\,meV at 0.16\,T, thereby suggesting an enhanced  $g$-factor.
The estimated $g$-factor enhancement is 17.5 times that of the bare value.
It is considered that a low electron density and high mobility makes it possible to reach a strong many-body interaction regime in which this type of strong enhancement in $g$-factor can be observed.}

\begin{document}
\maketitle

\section{Introduction}

Tunnel-coupled double-layer two-dimensional electron gas (2DEG) systems exhibit several interesting phenomena due to two internal degrees of freedom, spin and pseudospin (layer index).
A well-known phenomenon in single-layer systems, such as quantum Hall effect (QHE), reveals further richness in double-layer systems\,\cite{Ezawa_book,annurevEisenstein,Sawada,Kumada329,SL_PRL,SL_PRB,JIALi}. 
For example, a prediction of Kosterlitz-Thouless transition in association with the dissociation of pseudospin vortices (``meron''s) is discussed\,\cite{Yang,Moon,Lay,KT_PRB}. 
Interestingly, these topologically protected pseudospin quasiparticles are considered to be non-Abelian. 
Furthermore, recent theoretical studies on topological quantum computing explored double-layer QHE systems that can host numerous non-Abelian quasiparticles\,\cite{Alicea,Ardonne,Peterson,Vaezi,Geraedts,Calixto,Barkeshli_2012,Barkeshli_2014,Barkeshli,ZhuW}.
However, the role of spins in double-layer QHE systems is unclear\,\cite{Terasawa_sky,Kumada_Spin,Luin,TsudaPRBSU4}, because controlling spin and pseudospin degrees of freedom individually is difficult.
This difficulty in controlling the spin and pseudospin degrees of freedom hampers precise identification of QHE ground states and topological quasiparticles. 
Therefore, the development of a selective spin filtering technique is required for double-layer systems. 

For this purpose, using a quantum point contact (QPC) is a feasible technique\,\cite{Debray,Hitachi,Hashisaka_NatPhys,Zimmermann}. 
A previous double-layer QPC study\,\cite{bilayerQPC} suggests that the system has an excessive interaction regime, in which a strong potential gradient produces an enhanced spin-orbit interaction and effective screening, as a result of high mobility electrons with a very low density.
This strong interaction regime possibly leads to an enhanced Land\'{e}'s $g$-factor because the low electron density\,\cite{WangBerggrenR} and strong confinement\,\cite{Ivchenko} increases the electron-electron interaction, and thereby increases the exchange interaction.
Such a situation is preferable for manipulating spins using the Zeeman effect. 
In the aforementioned study\,\cite{bilayerQPC}, we used in-plane magnetic fields and observed a $g$-factor of twice that of the bare GaAs value. 
Enhancements in $g$-factor for in-plane fields are also observed in previous studies\,\cite{Thomas,Thomas_PRB1998,Martin_ApplPhysLett,Martin_PRB,Chen_PRBR}.
However, $g$-factor is not isotropic\,\cite{Ivchenko,Malinowski,Nefyodov_1,Nefyodov_2}.
Thus far, only a few conductance measurements in the presence of out-of-plane magnetic fields are conducted to date\,\cite{vanWees_PRB,Thomas_APL,Roessler,Martin_PRB,Baer,FangchaoLu,Nichele_2014,Overweg,Kraft}, and there is a paucity of reports on double-layer QPCs.
In the experiments above, enhancements in $g$-factor were reported that deserved  theoretical attention\,\cite{WangBerggrenR,Aryanpour,Vionnet,Klironomos_EPL,Bruus,Yang1D}.
Thus, we believe that further elaborated studies for double-layer GaAs/AlGaAs QPC systems in the presence of an out-of-plane magnetic field will provide valuable information on spin manipulation and spin filtering.

In this study, we fabricated a QPC in a double-layer 2DEG of GaAs/AlGaAs double quantum well (DQW) sample and examined a small out-of-plane magnetic field effect on it. 
To explore the possibility of manipulating electron spins, we determined the remaining basic spin-splitting properties of a double-layer QPC system by investigating the Zeeman gap based on the $g$-factor in the out-of-plane direction.
Owing to the magnetic confinement, subband edges (SBEs) are parabolically bended towards higher energy\,\cite{Glazman_JPhysCondMatt,Berggren}.
We observe a strong perpendicular field dependence of SBEs. 
Furthermore, the Zeeman gap splitting of SBEs clearly appears at 0.1\,T, and becomes 0.09\,meV at 0.16\,T.
From the magnetic field dependence of the Zeeman splitting, we derive an enhanced $g$-factor of 7.7 (17.5 times the bare value). 
We discuss the possible contribution of electron-electron interaction to the enhancement of the $g$-factor. 
The results are promising for spintronics and quantum computation.

The remainder of this paper is organized as follows. In Section\,\ref{experiment}, the sample structure and the experimental methods are described. In Section\,\ref{Results}, the experimental results and discussion are presented. Finally, brief concluding remarks are presented in Section\,\ref{conclusion}.

\section{Experimental Details}
\label{experiment}

\begin{figure}  
\includegraphics[width=1\linewidth]{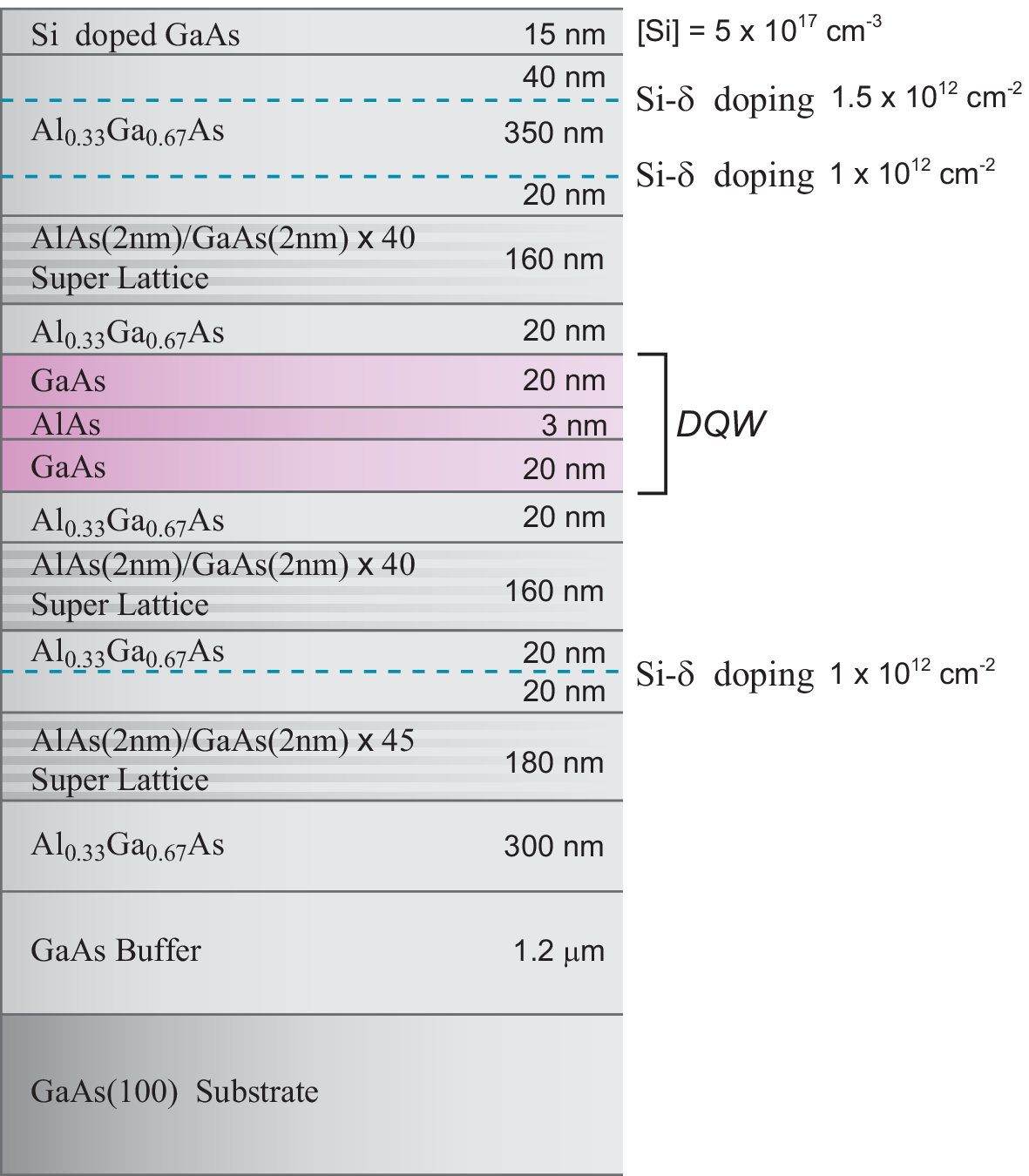}
\caption{(Color online)\,Schematic illustration of the DQW sample.}
\label{fig_structure}
\end{figure}

The sample used in the study is identical to the sample that was used in another literature\,\cite{bilayerQPC}. 
Figure \ref{fig_structure} shows a schematic illustration of the sample layer sequence.
The DQW heterostructure is grown by molecular beam epitaxy on the GaAs (100) surface in NTT Basic Research Laboratories.
Epitaxial layers with DQW (two 20-nm-wide GaAs quantum wells separated by a 3-nm-wide AlAs barrier layer) is located 605\,nm below the surface, and is doped from both sides via $1 \times 10^{12}$\,cm$^{-2}$ Si $\delta-$dopings that are 200\, nm away from each side of DQW. 
The electron density in the symmetric state corresponds to $0.64 \times 10^{11}$ cm$^{-2}$ and that in the anti-symmetric state corresponds to $0.56 \times 10^{11}$ cm$^{-2}$, with an energy gap of 0.29\,meV between them\,\cite{bilayerQPC}.
The low temperature electron mobility is approximately $2.5 \times 10^6 $\,cm$^2$/(Vs).
A standard Hall bar 
is fabricated with AuGe/Ni ohmic electrodes in contact with both layers.
A pair of split gates with a width of 500\,nm and a length of 100\,nm is fabricated at the center of the Hall bar via electron beam lithography technique.
The sample is mounted upside-down on the cold finger of the mixing chamber of a dilution refrigerator with a base temperature corresponding to 20\,mK.

\begin{figure}  
\includegraphics[width=1\linewidth]{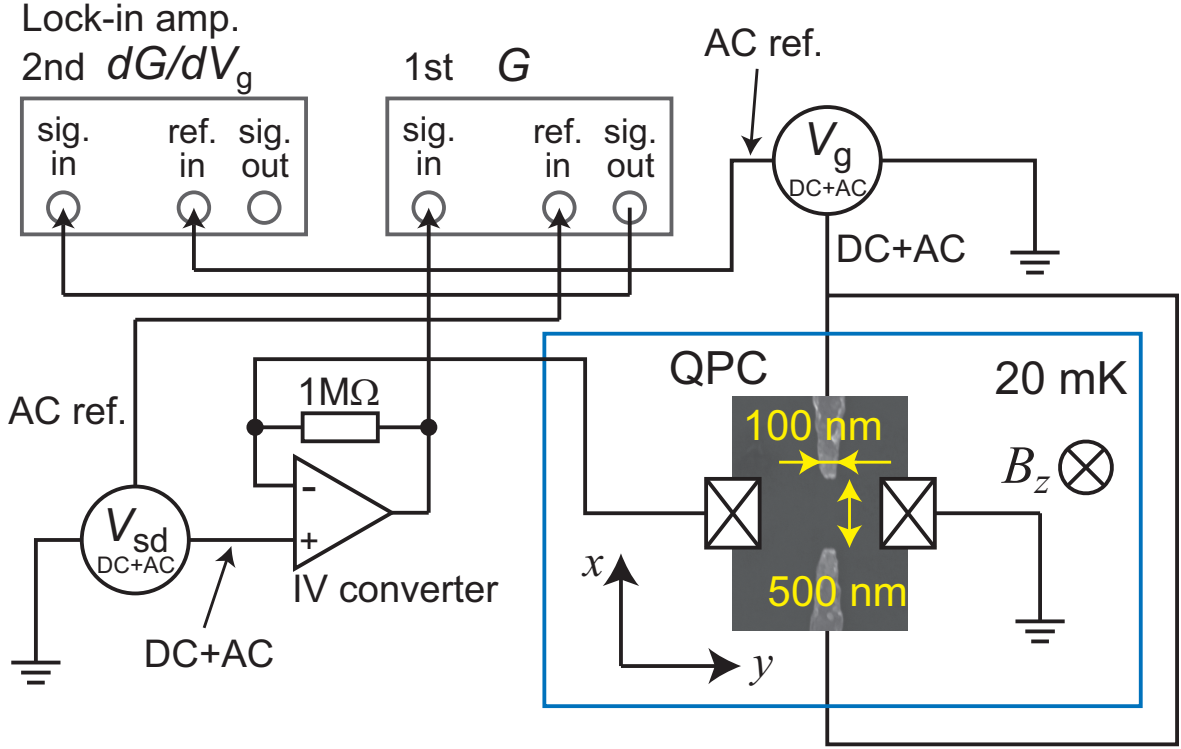}
\caption{(Color online)\,Scanning electron microscopy image of split gates (QPC) and schematic image of the measurement setup. Sample is placed upside down on the cold finger of the mixing chamber, and its drain is grounded. Magnetic field $B_z$ is applied perpendicular to the sample. }
\label{fig_sample}
\end{figure}

Figure \ref{fig_sample} shows a scanning electron microscopy image of the split gates and schematic image of the measurement.
Two-terminal differential conductance $G = dI_{\rm sd}/dV_{\rm sd}$ (where $I_{\rm sd}$ and $V_{\rm sd}$ denote the source-drain current and voltage, respectively) and transconductance $dG/dV_{\rm g}$ ($V_{\rm g}$ denotes the gate voltage applied to the split gates) are simultaneously measured using two lock-in amplifiers.
First, $G$ is measured via the first lock-in amplifier with a frequency of 387\,Hz and amplitude of $V_{\rm sd}^{\rm ac} =10$\,$\mu$V r.m.s., and a small AC gate modulation $V_{\rm g}^{\rm ac}=4$\,mV r.m.s. is simultaneously applied via the second lock-in amplifier with a frequency of 13\,Hz.
The output signal of the first lock-in amplifier (which includes the AC modulation signal from $V_{\rm g}^{\rm ac}$) is input to the second lock-in amplifier.
This method enables precise direct measurement of transconductance.
However, conductance slightly becomes noisy. Therefore, conductance changes that are not supported by concurrent transconductance changes can possibly be experimental noise.
A DC gate voltage $V_{\rm g}^{\rm dc}$ is also applied to the sample, and thus the total voltage applied to the split gate $V_{\rm g}$ is $V_{\rm g}=V_{\rm g}^{\rm dc} + V_{\rm g}^{\rm ac}$. 
Additionally, we apply a DC voltage to the source to cancel the voltage of the Seebeck effect and to induce a nonequilibrium bias.  
Hence, the total voltage applied to the source $V_{\rm sd}$ is $V_{\rm sd} =  V_{\rm sd}^{\rm ac} + V_{\rm sd}^{\rm dc}$, where $V_{\rm sd}^{\rm dc}$ denotes the total DC voltage applied to the sample. 
In the graphs and image plots, we ignore the AC component of $V_{\rm g}$ and $V_{\rm sd}$ for practical reasons.
The $x$, $y$, and $z$-directions are as follows: $x$-direction is perpendicular to the current and in-plane to 2DEG (see also Fig. 1); $y$-direction is parallel along the current and in-plane to 2DEG; and $z$-direction is perpendicular to 2DEG.  
A $z$-directional magnetic field $B_z$ is applied using a superconductor (vector) magnet, with the maximum field of $B_z=8$\,T.

\section{Results and Discussion}
\label{Results}

\begin{figure} 
\includegraphics[width=0.8\linewidth]{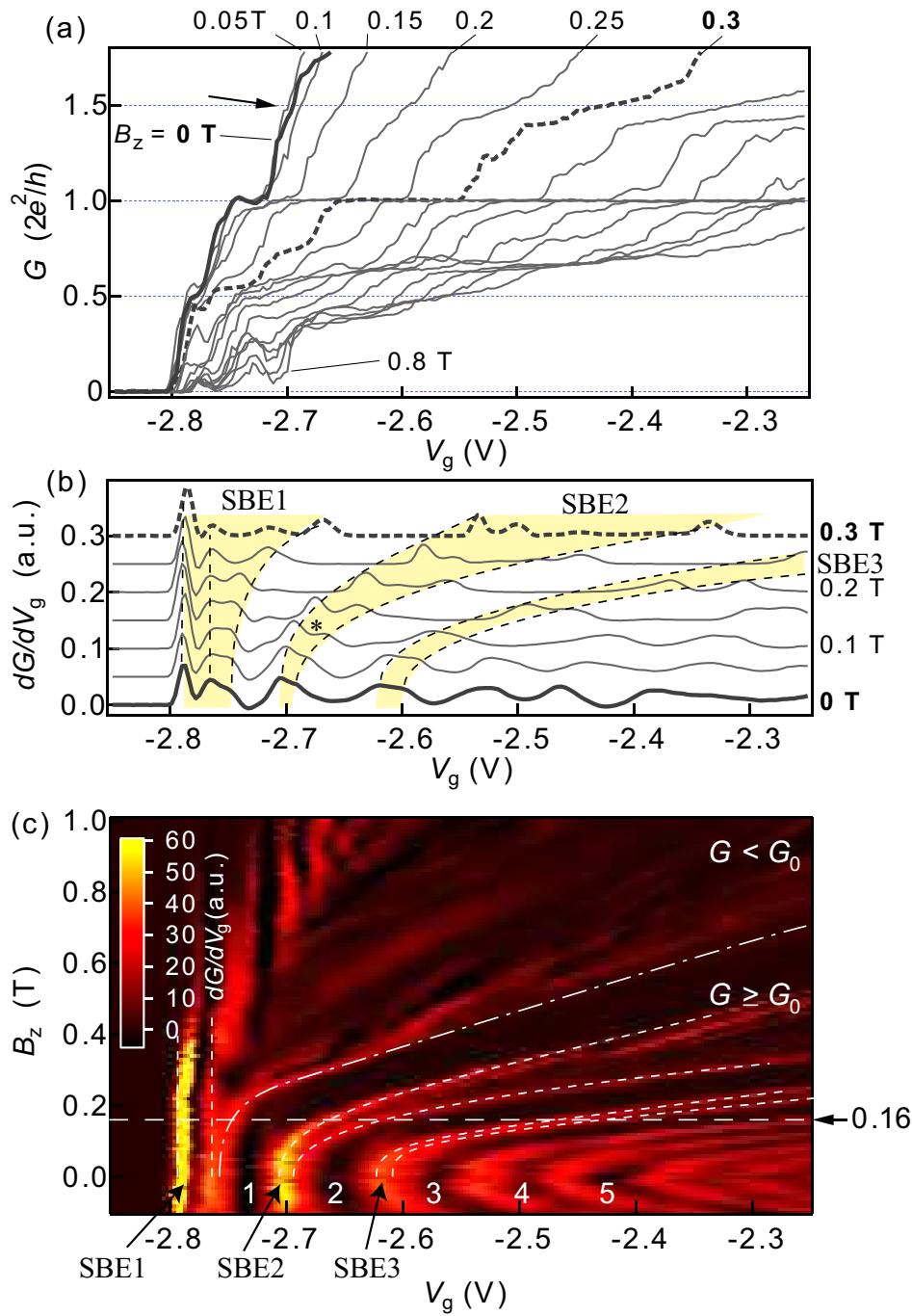}	
\caption{(Color online)\,(a) $G$ in the unit of $2e^2/h (=G_0)$ as a function of $V_{\rm g}$ for several $B_z$ values from 0 (bold line) to 0.8\,T by 0.05\,T step. The dashed line indicates $G$ for $B_z = 0.3$\,T. The black arrow indicates a signature of $1.5 G_0$ plateau. (b) $dG/dV_{\rm g}$ as a function of $V_{\rm g}$ for $B_z = 0$ (bold line) to 0.3\,T (dashed line) by 0.05\,T step. Each trace is offset for clarity. $\ast$ mark indicates the minimum of $1.5G_0$ plateau. The dotted lines are visual guide lines. (c) Image plot of $dG/dV_{\rm g}$ as a function of $B_z$ and $V_{\rm g}$ for $V_{\rm sd}=0$\,V. The numerical numbers express the corresponding plateaus in the unit of $G_0$. The dash-dotted line divides $G \geq G_0$ and $G < G_0$ regions. The dotted lines indicate SBE peaks of interest in Fig.\,\ref{fig_dEz}. The dashed line indicates the corresponding $B_z$ value for the next figure, Fig.\,\ref{fig_dGdVg_Bz}.}
\label{fig_Bz_G_dGdVg}
\end{figure}

Figure \ref{fig_Bz_G_dGdVg} (a) shows $G$ in the unit of $G_0=2e^2/h$ ($e$ is elementary charge, and $h$ is Planck's constant) as a function of $V_{\rm g}$ for different $B_z$ values ranging from 0 to 0.8\,T (0.05\,T step).
Fundamental conductance properties are similar to those of the single-layer GaAs 2DEGs\,\cite{vanWees,Wharam}.
The energy level of electrons becomes quantized due to nanometer-scale lateral confinement at QPC, and thus, the conductance is described by the Landauer-B\"{u}ttiker model\,\cite{ButtikerImryLandauerPinhas,Buttiker}.
A clear conductance plateau at $G \simeq 0.5 G_0$ is observed at $B_z=0$\,T (bold line).
In addition, a plateau at $G \simeq 1.5 G_0$ gradually develops as $B_z$ increases.
As subsequently demonstrated, the $1.5G_0$ plateau pertains to Zeeman splitting, and thus we focus on its subsequent change.
When we increase $B_z$,  the $G_0$ plateau shifts for larger $V_{\rm g}$ values and the plateau region is extended.
Further, conductance exhibits many small plateau-like features for $G < G_0$.

These features are also evident in the measurement of $dG/dV_{\rm g}$.  
Figure \ref{fig_Bz_G_dGdVg} (b) shows the $dG/dV_{\rm g}$ profile as a function of $V_{\rm g}$ for $B_z = 0$ to 0.30\,T by 0.05\,T. 
In the $dG/dV_{\rm g}$ plot, the plateau in $G$ and the crossing of SBEs correspond to minima and maxima, respectively.
With respect to $B_z = 0$\,T, we observed two peaks for the first integer SBE (SBE1), which soon resolved into three peaks (see ref.\,\cite{bilayerQPC}), and a broad peak for the second integer SBE (SBE2) and the third integer SBE (SBE3).
These peaks were resolved into two Gaussian peaks as shown in Fig.\,\ref{fig_dEz} (a).
Specifically, a small minimum appears in the $dG/dV_{\rm g}$ profile for $G = 1.5 G_0$ (indicated by $\ast$) at $B_z =0.1$\,T, thereby indicating spin splitting due to the opening of the Zeeman gap.

Figure \ref{fig_Bz_G_dGdVg} (c) shows an image plot of $dG/dV_{\rm g}$ as a function of $V_{\rm g}$ and $B_z$.
SBEs show a rapid increase relative to $V_{\rm g}$ as $B_z$ increases.
Apparently, SBE2 and SBE3 split into the two main peak lines indicated by dotted lines. 
As shown later in Fig.\,\ref{fig_dEz} (a), the broad peak consists of two smaller peaks at $B_z =0$, which indicates the existence of other spin-splitting contributions to this system.
The $dG/dV_{\rm g}$ bifurcation that corresponds to the $1.5G_0$ plateau in Fig.\,\ref{fig_Bz_G_dGdVg} (a) is clearly observed near a small field of $B_z \geq 0.1$\,T. 
This result should be noted because previous experiments such as \cite{Nichele_2014} require $B_z =1$\,T to split the second integer SBE, although the material used in the reference is different, namely, a GaAs heavy hole system that is considered to yield a larger Zeeman splitting.
The parabolic dependence of the SBEs is typically described as an additional effective confinement due to the cyclotron motion\,\cite{Berggren}, that is, $\frac{m^\ast}{2} \omega^2x^2$,
where $\omega^2 = \omega_0^2 + \omega_c^2 \, (\omega_c = eB_z/m^\ast)$, at the center of the QPC region.
In this system, a strong potential gradient along the $z$ direction is expected, and this gradient in the potential causes electrons to populate in one layer (back layer) of the DQW\,\cite{bilayerQPC}.
Hence, the low electron density of this sample ($\sim 0.6 \times 10^{11}$ cm$^{-2}$ per layer) accelerates the depopulation from the higher energy in the presence of a magnetic confinement.
Consequently, this sample embodies one of the lowest density regime in the QPC region in which a strong electron-electron interaction is expected.

However, in the $G<G_0$ region, as $B_z$ increases, the peak lines that belong to SBE1 show complicated bifurcations.
These lines have features that are not easily associated with the spin-resolved SBE lines.
As discussed later, these features are probably attributable to either Fabry-P\'{e}rot resonances\,\cite{CGSmith,Baer} or transmission resonances\,\cite{Tekman}, or to spin-dependent transmissions due to many-body interactions.
Although these observations are interesting and may broaden our understanding of a previous study\,\cite{Baer}, 
it is difficult to discuss the Zeeman splitting of SBE1 based on these lines. 
Therefore, we focus on the SBE2 and SBE3 peaks.

\begin{figure} 
\includegraphics[width=0.7\linewidth]{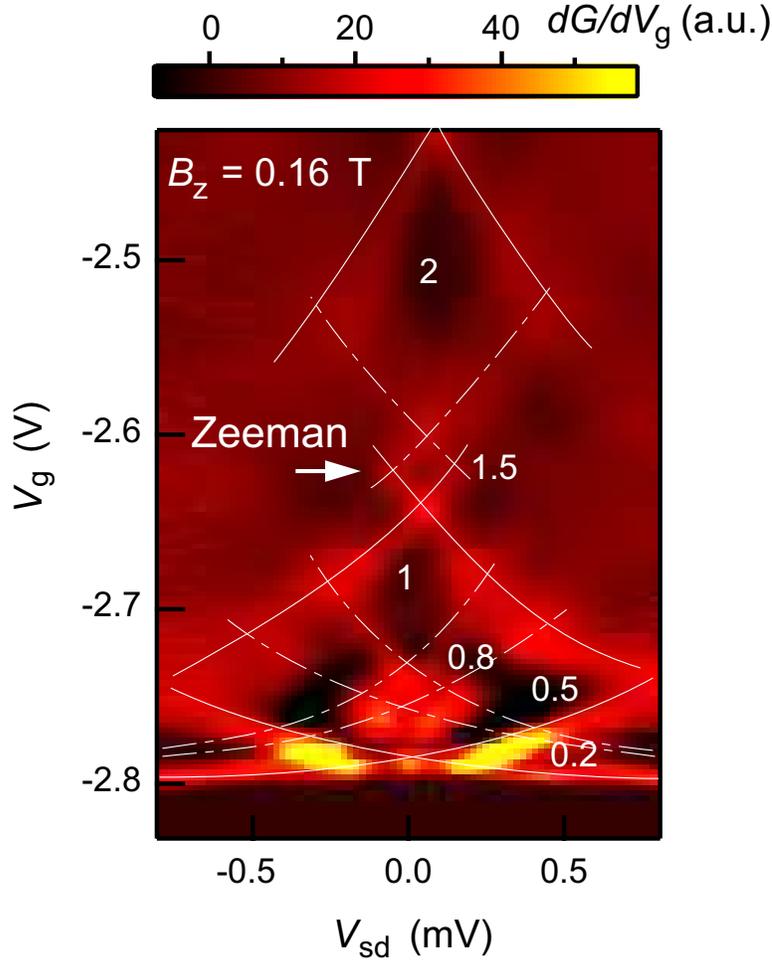}	
\caption{\label{fig_dGdVg_Bz}(Color online)\,Image plot of $dG/dV_{\rm g}$ as a function of  $V_{\rm sd}$ and $V_{\rm g}$ at $B_z=0.16$\,T. The numerals on the image plot represent approximate conductance values in $G_0$. The white solid lines indicate primary SBE lines; further splitting is indicated by dash-dotted  lines, which were drawn based on $dG/dV_{\rm g}$ maxima. }
\end{figure}

We use the non-equilibrium bias effect to convert the peak separation in $V_{\rm g}$ into Zeeman energy. 
Figure \ref{fig_dGdVg_Bz} shows $dG/dV_{\rm g}$ as a function of $V_{\rm sd}$ and $V_{\rm g}$ for $B_z = 0.16$\,T.
We connect the maxima (SBE) in $dG/dV_{\rm g}$ and draw solid lines for integer series and dash-dotted lines for split SBEs.
As shown in Fig.\,\ref{fig_Bz_G_dGdVg} (b), we observe three split SBE lines for $G \leq G_0$.
Importantly, a small diamond that corresponds to the $1.5 G_0$ plateau appears due to the increase in Zeeman energy (indicated by the white arrow). 
From this diamond pattern, the estimated magnitude of Zeeman splitting, along with the contributions of other spin splitting factors, corresponds to 0.09\,meV.  
The bare Zeeman energy $\epsilon_{\rm Z} = |g|\mu_{\rm B} B$ with $|g| = 0.44$, is calculated as $\sim 0.004$\,meV for $B_z=0.16$\,T, where $g$ denotes the Land\'{e} $g$-factor, and $\mu_{\rm B}$ denotes the Bohr magneton.

Subsequently, we extract the Zeeman splitting peak positions and estimate the enhancement in $g$-factor.
Figure \ref{fig_dEz} (a) shows an example of two-peak Gaussian curve fit for two convolved peaks.
Then the extracted peak positions (approximately, the broken lines in Fig.\,\ref{fig_Bz_G_dGdVg}) are converted into energies using the relationship between the energy gap and the diamond width at $B_z =0.16$\,T. 
In addition, it is necessary to consider the ``lever-arm'' correction to compare diamonds in different $V_{\rm g}$ value regions (Appendix \ref{LeverArm}). 
The result is shown in Figure \ref{fig_dEz} (b) that depicts energy gaps $\Delta E_{\rm Z}$ for SBE2 and SBE3 as a function of $B_z$.  
As shown in the figure, $\Delta E_{\rm Z}$ for both SBEs remain finite at $B_z=0$\,T.
With respect to  the lower $B_z$ region ($B_z \lesssim 0.17$\,T), if we express effective Zeeman energy $E_{\rm z}$ in the following form:
\begin{equation}
E_{\rm z} = \sqrt{E_0^2 + (g^\ast \mu_{\rm B} B_z )^2},       \label{Hz}
\end{equation}
where $E_0$ represents the spin-splitting contributions of the effective magnetic fields other than the $z$-directional applied field; this includes the contribution of the spin-orbit interaction, strong confinement, and electron-electron interaction, which are presumably oriented in the $x$-direction\,\cite{bilayerQPC}. 
$g^\ast$ denotes the $g$-factor in the $z$-direction.
Then, the fit using the above equation (the solid lines in Fig.\,\ref{fig_dEz} (b)) yields $E_0 = 0.060 \pm0.002$\,meV and $g^\ast \mu_{\rm B} = 0.45 \pm 0.02$\,meV/T for SBE2, and $E_0 = 0.050 \pm 0.002$\,meV and $g^\ast \mu_{\rm B} = 0.17 \pm0.03$\,meV/T for SBE3.
The magnitude of the enhanced $g$-factor $g^\ast$ extracted from the fit for SBE2 is $g^\ast \approx 7.7$, which is approximately 17.5 times the bare value, and $g^\ast \approx 2.9$ ($\approx 6.5$ times the bare value) for SBE3. 
These values are greater than the previously reported enhancement values in GaAs 2DEG systems ($0.4 \lesssim g^\ast \lesssim 1.3$ in \cite{Thomas} and $3.8 \lesssim g^\ast \lesssim 4.4$ in \cite{Roessler}) and GaAs two-dimensional hole systems ($3 \lesssim g^\ast \lesssim 7.2$ in \cite{Nichele_2014}).
As previously reported\,\cite{Thomas,Thomas_PRB1998,Martin_ApplPhysLett,Martin_PRB,Chen_PRBR}, the $g^\ast$ value decreases as the subband index increases, because the 1D confinement becomes stronger for lower subbands.
In addition, the difference in $g^\ast$ between SBE2 and SBE3 is probably attributed to the density difference\,\cite{WangBerggrenR}.
With respect to the higher $B_z$ region ($B_z > 0.17$\,T), the data deviate from the fit, which indicates a further enhancement in the $g^\ast$ value. As we can approximate $dE_{\rm Z}/dB_z \simeq g^\ast\mu_{\rm B}$ for the higher $B_z$ region, we obtained a considerably higher value of $g^\ast \mu_{\rm B} = 0.96 \pm0.08$\,meV/T from the slope (the dashed line in Fig.\,\ref{fig_dEz} (b)).
Such large values were obtained using the density functional theory\,\cite{WangBerggrenR}.
For this region, the parabolic confinement due to $B_z$ contributes to the apparent enhancement in the splitting\,\cite{WangBerggrenR,Vionnet}. 

%
\begin{figure} 
\includegraphics[width=1\linewidth]{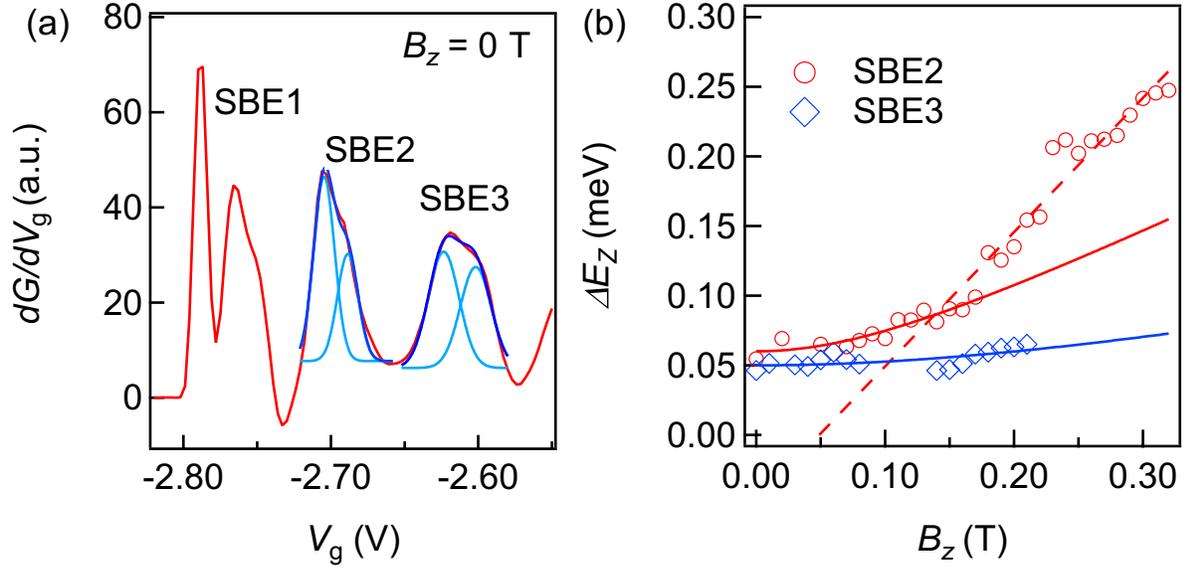}	
\caption{(Color online)\,(a) $dG/dV_{\rm g}$ as a function of $V_{\rm g}$ at $B_z=0$\,T, showing an example of extracted peak positions from the fitting of two-peak Gaussian curves for overlapping peaks.  (b) Lever-arm-corrected $\Delta E_{\rm z}$ as a function of $B_z$ for SBE2 and SBE3. }
\label{fig_dEz}
\end{figure}

We consider that the electron-electron many-body interaction is the underlying cause of the enhanced $g$-factor. 
In an earlier experiment, Thomas {\it et al}.\,\cite{Thomas_SpinProperties} observed 0.5 plateau that suggests a spin polarized state in zero magnetic field, in which the lower density enhances the many-body interaction.
Considering that the low electron density and high mobility, the sample used in this experiment offers a unique opportunity to realize a state with significant many-body interaction effects.
Nuttinck {\it et al}.\,\cite{Nuttinck} also indicated a 0.5 plateau at zero magnetic field along with a 0.7 shoulder.
Considering that the present sample has the quality that is comparable to the sample used by Nuttinck {\it et al}., a similar regime in terms of impurity effect is realized in the experiment.
Subsequently, the large Zeeman gap and 0.5 plateau should be attributed to a large electron-electron exchange interaction, as proposed in Ref. \cite{Thomas,Thomas_SpinProperties,Nuttinck} and theoretically indicated in Ref. \cite{WangBerggrenR,Aryanpour,Vionnet,Klironomos_EPL}.
Along with the exchange interaction, the Zeeman gap can have the contribution of the spin-orbit interaction, as speculated in our previous work\,\cite{bilayerQPC}.
However, as Ref.\,\cite{WangBerggrenR} shows that a large $g$-factor enhancement is accountable from the exchange contribution, we infer that the spin-orbit interaction contributes less to the $z$-directional $g$-factor enhancement than the contribution by exchange interaction.
It is immediately observed that the enhanced value is larger than that of a single-layer system ($g^\ast \lesssim 1.3$ in \cite{Thomas} ).
Regarding this result, the DQW constriction may constructively affect the $g$-factor enhancement; however, no theoretical studies to support this idea exist to date.
Additionally, we observed Zeeman splitting approximately corresponding to 0.09\,meV in the presence of an in-plane and perpendicular-to-current magnetic field $B_x$ of 2.0\,T\,\cite{bilayerQPC}. 
In this case, the value is approximately twice the bare Zeeman splitting, and thus the difference between $x$ and $z$-directions is evident, as observed in two-dimensional hole systems\,\cite{Nichele_2014,Srinivasan}. 
As discussed in Ref.\,\cite{Ivchenko}, this remarkable difference in the $g$-factor is attributed to the difference in the electric confinement between the $x$- and $z$-directions. 

Further, the strong many-body interaction may affect the conductance.
As shown in Fig.\,\ref{fig_Bz_G_dGdVg} (a) and (c), we observe the conductance plateaus for $G<G_0$ and $dG/dV_{\rm g}$ peak lines for $B_z \geq 0.3$\,T [for further details, see Appendix \ref{DataBz0.3}].
These features may relate to the spin and pseudospin degrees of freedom in double-layer systems; however, the numerical simulation results suggest the deficient involvement of the pseudospin degree of freedom\,\cite{bilayerQPC}. 
Theoretically, the possibility of forming a quasi-bound state of electrons in the strong interaction regime was discussed\,\cite{Rejec,Rejec_CondMatt,Flambaum,Rejec_PRB,Shelykh,Meir}.
If this is the case, the transmission coefficient depends on the spin configuration of the quasi-bound state whether the quasi-bound state is singlet or triplet.
Further investigation is required to clarify the relationship between the conductance and spin configurations.

\section{Concluding Remarks}
\label{conclusion}

In conclusion, the results of the study indicate a rapid and strong SBE dependence on $B_z$ in a double-layer QPC. 
The results reveal that the Zeeman gap opening begins to appear at $B_z=0.10$\,T.
It is important to note that the estimated enhancement in the $g$-factor is 17.5 times the bare value.
We attribute the $g$-factor enhancement to a strong electron-electron interaction due to low electron density and high mobility.
We believe that the results are profitable to manipulate spins in double-layer systems.

\begin{acknowledgment}

The authors express their gratitude to K. Muraki and T. Saku of the NTT basic research laboratories and A. Sawada for providing a high-mobility sample.
This study was supported by the JSPS KAKENHI (JP15K17680, JP15H05854, JP18H01815, JP19H05826, JP19H00656).
\end{acknowledgment}

\appendix

\section{``Lever Arm" Corrections}
\label{LeverArm}

\begin{figure}
\includegraphics[width=0.8\linewidth]{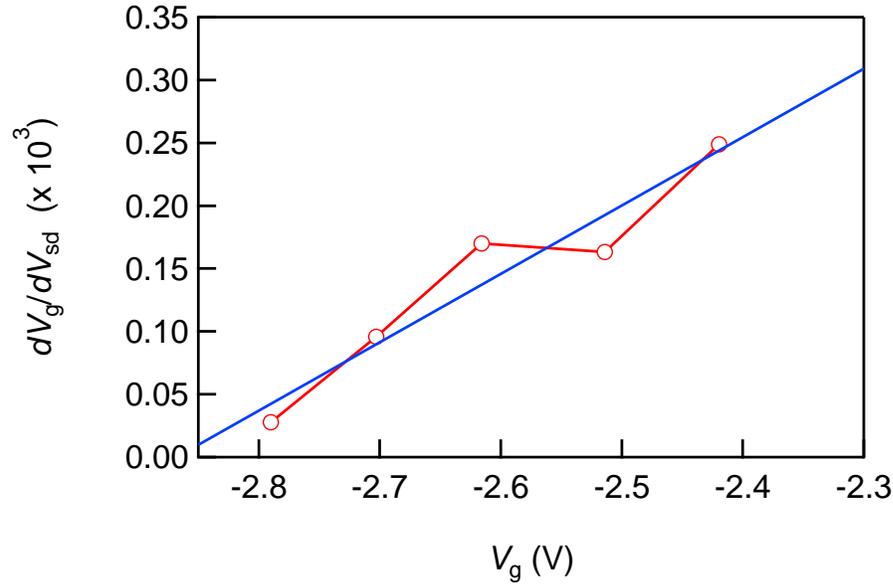}	
\caption{\label{fig_LeverArm}(Color online)\,Slope correction of the subband edge lines $dV_{\rm g}/dV_{\rm sd}$ as a function of $V_{\rm g}$. }
\end{figure}

As shown in a certain reference\,\cite{Roessler}, the $V_{\rm g}$ dependence between subband edges (the difference between $dG/dV_{\rm g}$ maxima) can be converted into energy gaps by using the corresponding $V_{\rm sd}$ differences.
However, the conversion coefficient $A$ (termed as the ``lever arm'') varies depending on $V_{\rm g}$, and thus it needs corrections as a function of $V_{\rm g}$.   
We deduced this lever arm correction in $A$ from the slopes of subband edge lines $dV^n_{\rm g}/dV_{\rm sd}$.
Specifically, $V^n_{\rm g}$ denotes the $n$-th integer SBE line.
Figure \ref{fig_LeverArm} shows deduced $dV^n_{\rm g}/dV_{\rm sd}$ as a function of $V_{\rm g}$ at 0\,T.
The slopes are derived from $-0.5 < V_{\rm sd} < -0.1$ mV region.
From a line fit, we obtain $dV_{\rm g}/dV_{\rm sd}  = \alpha_0 + \alpha_1 V_{\rm g}$, and subsequently the lever arm coefficient as follows:
\begin{equation}
A(V_{\rm g}) = \frac{A_0}{\alpha_0 + \alpha_1 V_{\rm g} },
\end{equation}
where $\alpha_0 = 1.56$ and $\alpha_1 = 0.54$\,V$^{-1}$.
Thus, the Zeeman gap $\Delta E_{\rm Z}$ is modified as follows:
\begin{equation}
\Delta E_{\rm Z} = A \cdot \Delta P,
\end{equation}
where $\Delta P$ denotes observed $V_{\rm g}$ gaps (between two $dG/dV_{\rm g}$ peaks in $V_{\rm g}$).

\section{Conductance at $B_z =0.30$\,T} 
\label{DataBz0.3}

\begin{figure}
\includegraphics[width=1\linewidth]{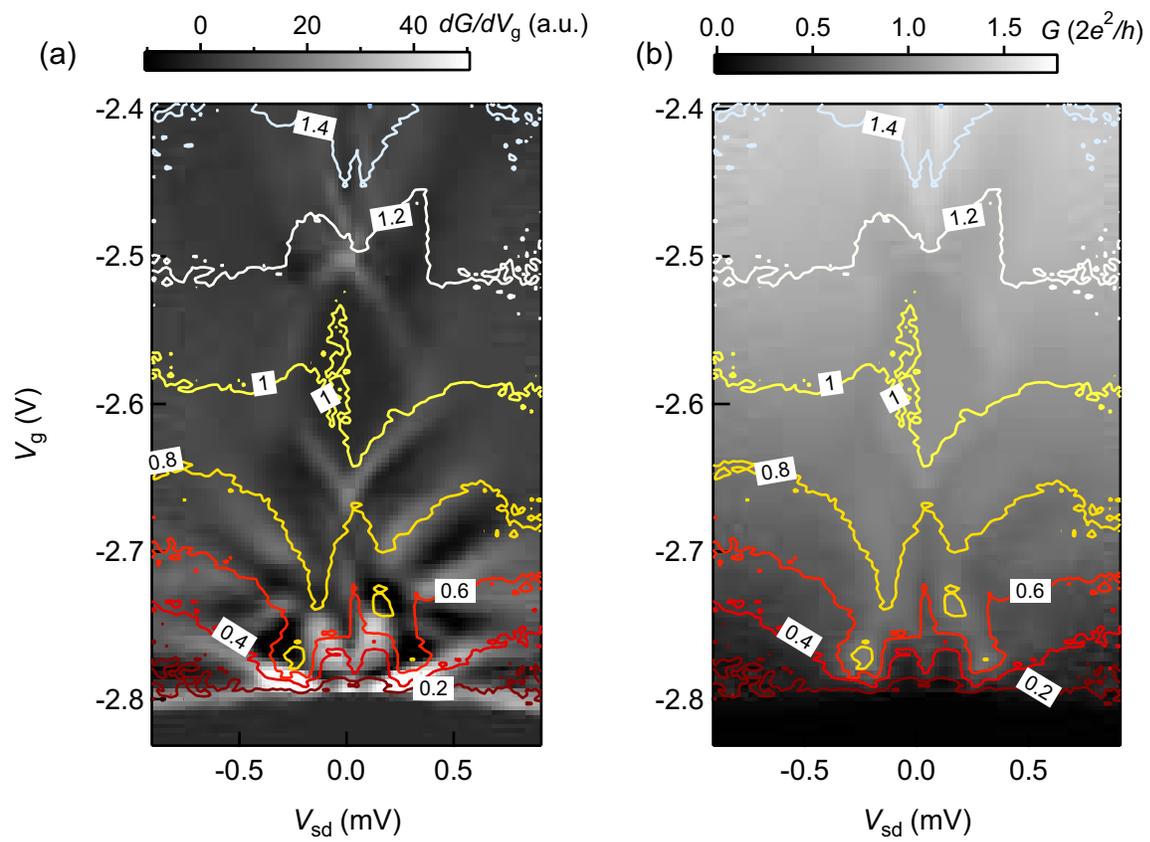}
\caption{\label{fig_G_dGdVg_0p3T}(Color online)\,(a) and (b) Image plot of $dG/dV_{\rm g}$ and $G$ as a function of $V_{\rm sd}$ and $V_{\rm g}$ with incorporated equi-conductance lines. }
\end{figure}

For further information on conductance change at $B_z=0.3$\,T, we append two image plots.
Figures \ref{fig_G_dGdVg_0p3T} (a) and (b) show image plots of $dG/dV_{\rm g}$ and $G$ at $B_z=0.3$\,T as a function of $V_{\rm sd}$ and $V_{\rm g}$ where equi-conductance contour lines are incorporated. 
The black-and-white scale in (a) starts from -10, and thus dark black regions indicate decreases in the conductance.
As shown in the figure, $G$ increases by approximately $0.2G_0$ at each $dG/dV_{\rm g}$ maxima (SBE) line for the $G < G_0$ region.

%


\end{document}